# Experimental Demonstration and Transformation Mechanism of Quenchable Two-dimensional Diamond

Jiayin Li,[1,2] Guoshuai Du,[1,3] Lili Zhao,[4] Wuxiao Han,[1,3] Jiaxin Ming,[1,2] Shang Chen,[1] Pengcheng Zhao,[1,3] Lu Bai,[1,2] Jiaohui Yan,[1] Yubing Du,[1,3] Jiajia Feng,[5] Hongliang Dong,[5] Ke Jin,[1] Weigao Xu,[4] Bin Chen,[5] Jianguo Zhang,[6,*] and Yabin Chen[1,7,*]

[1]*School of Interdisciplinary Science, Beijing Institute of Technology, Beijing 100081, China*

[2]*School of Chemistry and Chemical Engineering, Beijing Institute of Technology, Beijing 100081, China*

[3]*School of Aerospace Engineering, Beijing Institute of Technology, Beijing 100081, China*

[4]*State Key Laboratory of Coordination Chemistry, Key Laboratory of Mesoscopic Chemistry of MOE, School of Chemistry and Chemical Engineering, Nanjing University, Nanjing 210023, China*

[5]*Center for High Pressure Science and Technology Advanced Research, Shanghai 201203, China*

[6]*State Key Laboratory of Explosion Science and Technology, Beijing Institute of Technology, Beijing, 100081, China*

[7]*Beijing Institute of Technology, Zhuhai Beijing Institute of Technology (BIT), Zhuhai, 519088, China*

[*]Correspondence and requests for materials should be addressed to: zjgbit@bit.edu.cn (J.Z.), or chyb0422@bit.edu.cn (Y.C.)

ABSTRACT

Two-dimensional (2D) diamond has aroused tremendous interest in nanoelectronics and optoelectronics, owing to its superior properties and flexible characteristics compared to bulk diamond. Despite significant efforts, great challenges lie in the experimental synthesis and transformation conditions of 2D diamond. Herein, we have demonstrated the experimental preparation of high quality 2D diamond with controlled thickness and distinguished properties, realized by laser-heating few-layer graphene in diamond anvil cell. The quenched 2D diamond exhibited narrow $T_{2g}$ Raman peak (linewidth ~3.6 cm$^{-1}$) and intense photoluminescence of SiV$^-$ (linewidth ~6.1 nm) and NV$^0$ centers. In terms of transformation mechanism, atomic structures of hybrid phase interfaces suggested that the intermediate rhombohedral phase subtly mediate hexagonal graphite to cubic diamond transition. Furthermore, the tunable optical bandgap and thermal stability of 2D diamond sensitively depend on its $sp^3$ concentration. We believe our results can shed light on the structural design and preparation of many carbon allotropes and further uncover the underlying transition mechanism.

**INTRODUCTION**

Two-dimensional (2D) diamond has attracted considerable interest due to its atomically thin structure in $sp^3$-hybridization and unique quantum confinement effect.[1-3] Bulk diamond is well known for its exceptional hardness, high thermal conductivity, and chemical inertness. However, it exhibits relatively low fracture toughness and poor electrical conductivity compared to many metallic and ceramic materials.[4-7] Various strategies have been developed to optimize its physical properties, such as structural modification and chemical doping.[8-15] The synthesized amorphous carbon, which exhibited a short-range ordered diamond-like structure, demonstrated its tunable electronic bandgaps although its thermal conductivity is significantly reduced compared to crystalline counterpart.[9] Nanotwinned diamond was developed with their simultaneous enhancement of hardness and toughness following the Hall-Petch relation.[11,12] Alternatively, thickness, as a novel degree of freedom, is rarely considered to regulate the diamond structure and further extend its applications, and the related research remains unmatured.[16]

It is predicted that 2D diamond can inherit the superior properties of bulk diamond and exhibit novel physicochemical properties, raised from its 2D characteristic.[17] Chernozatonskii *et al.* proposed the $C_2H$ structures as "diamane", by absorbing hydrogen atoms on bilayer graphene.[18] Theoretical simulations underline that chemical functionalization and stacking sequence effectively modulate bandgap and thermal conductivity of 2D diamond.[19,20] Extensive efforts have been paid to synthesize 2D diamond in experiments, such as the direct phase transformation from few-layer graphene (FLG) to diamondol[21], diamane[22-24], diamene[25-27], and diamondene[28-30]. Hydrogenated and fluoride 2D diamond have been achieved by chemical adsorption of FLG under moderate conditions, while this approach usually produces the incomplete internal transformation.[22-24] Moreover, bilayer graphene on SiC reversibly transformed into diamond-like film upon tip compression, which was severely hindered by the stacking sequence of FLG films.[26] Thin hexagonal diamond (HD) synthesized from FLG and its layer-dependent transition pressure was elucidated, and this metastable phase can be unfortunately preserved to ~ 1.0 GPa during decompression.[27] Biaxial strain approach was explored to play an important role in the reversible phase transition from multilayer graphene (13 ~ 94 nm thick) to diamondene on different substrates.[30] In theory, high pressure and high temperature (HPHT) are supposed to irreversibly transform FLG into the quenchable 2D diamond,[31] which remains to be further confirmed experimentally.

To understand the daunting transition mechanism, numerous pathways have been proposed to illustrate the transformation at the atomic scale, such as the concerted mechanism and nucleation assumption.[32-34] Theoretical simulation predicts that hexagonal graphite (HG) lattice undergoes slipping, followed by puckering to form cubic diamond (CD) and HD.[34] Rhombohedral graphite (RG), as an intermediate phase, may play a key role for HG to CD transformation.[35] Meanwhile, considerable efforts have been paid to study the heterophase structures between diamond and graphite, including the van der Waals interaction-governed and covalently bonded diaphite with the fancy interfacial motifs.[36-38] Compared to bulk diamond, significant gap still remains to elaborate the transition mechanisms in 2D diamond, both experimentally and theoretically. Theoretical results indicate that nucleation-and-growth mechanism, structural defects, and interlayer sliding may be essential to 2D diamond formation.[30,39,40]

Despite tremendous efforts invested, the controlled synthesis and phase transition mechanism of 2D diamond remain persistently challenging. Here, we report a novel strategy for 2D diamond synthesis with controlled thickness extending from bilayer graphene to several hundred nanometers, by laser-heating metal-supported graphene precursor under high pressure. Lattice structures of recovered samples together with the hybrid interfaces evidently confirm the mediation of RG in HG to CD transition and crystallographic orientational relationships following $(002)_{HG} // (003)_{RG} // (111)_{CD}$ and $(002)_{HG} // (100)_{HD}$. Moreover, our results reveal that the prepared ultrathin diamond exhibits prominent thermal stability, and its sizable bandgaps vary from 1.4 to 1.9 eV, originated from its obvious $sp^3$ content-dependence. This work can provide great insights into the structural control and long-stand puzzle in transition mechanism of 2D diamond.

## RESULTS

### 2D diamond synthesis under HPHT

First, we performed HPHT preparations of 2D diamond from graphene precursors with various thicknesses, ranging from bilayer to several hundred nanometers (Supplementary Figures 1-3 and Supplementary Table 1). Regarding to its low absorbance of graphene (2.3% for monolayer limit),[41] we established a strategic approach to load FLG into diamond anvil cell (DAC), followed by laser heating to achieve HPHT conditions. The polished metallic Re foil was utilized as both graphene substrate and laser absorber to endure high temperature (Figure 1 and Supplementary Figure 1), due to its high melting point (3450 K at 1 atm), unique mechanical modulus and carbide-free characteristic.[42,43] The prepared graphene/Re was sealed in laser-heated DAC using KBr as the pressure transmission medium (PTM) and thermal insulator, as illustrated in Figure 1a. Both HD and CD phases are expected after HPHT treatment, and the latter one usually undergoes a lower-barrier pathway[35] in Figure 1b. Figure 1c-f show the representative optical evolutions of a trilayer (3L) graphene during the entire experimental process. The FLG on Re substrate exhibited weak optical contrast resulting from its low absorbance. The morphology of FLG can be precisely traced from precursor preparation through HPHT treatment, even after being quenched to ambient condition from 32.3 GPa and 2924 K, as indicated in Figure 1c-f.

Raman spectrum before and after HPHT treatment were measured to verify the transition from FLG to 2D diamond, as shown in Figure 1g. Raman spectrum of pristine 3L graphene before compression displays the prominent G peak at 1580 cm$^{-1}$ and a complete absence of the D band, confirming that our optimized transfer protocol preserves the defect-free nature of graphene precursor. The quenched 3L sample from 32.3 GPa and 2924 K exhibits an evident Raman $T_{2g}$ peak around 1332 cm$^{-1}$, which is uniquely originated from the lattice structure of *sp*$^3$ carbon, i.e., diamond structure (the detailed Raman characterization will be discussed later).

The uniform and ultrathin nature of the quenched 2D diamond was confirmed by high-resolution transmission electron microscopy (HRTEM) and energy-dispersive X-ray spectroscopy (EDS) characterizations. HRTEM results (Figure 1h) reveal a continuous carbon layer with a uniform thickness of approximately 1.0 nm, sandwiched between the gold coating and the Re substrate as confirmed by the EDS mapping results in Figure 1i. The thickness of the quenched 2D diamond derived from 3L graphene is comparable to the theoretical thickness (9 Å) of a double-sided hydrogenated 3L diamane.[44]

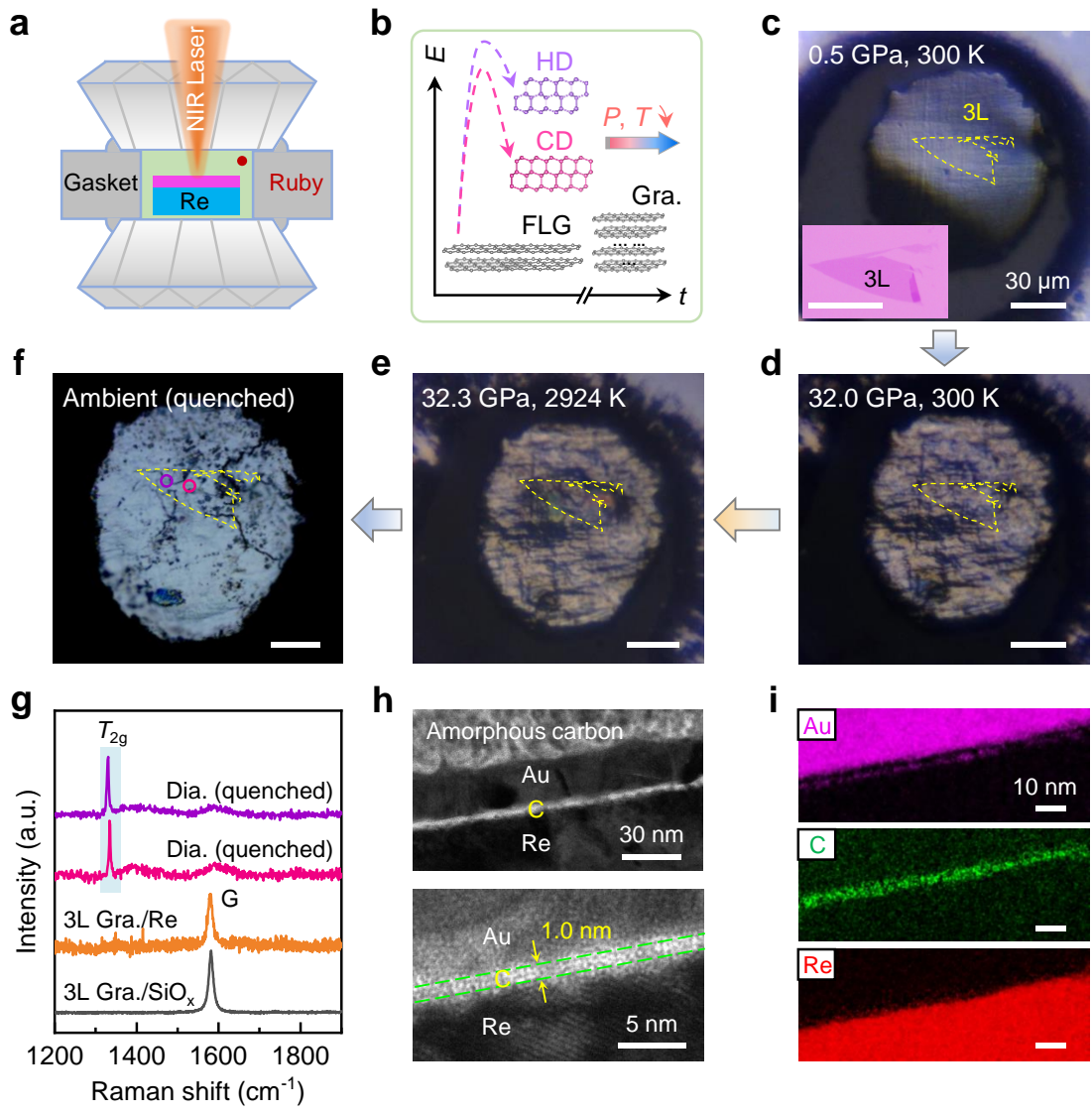

**Figure 1. Experimental synthesis of 2D diamond under HPHT. a**, Schematic strategy to prepare 2D diamond from graphene in laser-heated DAC. Re foil (light blue), polished down to ~ 10 μm with surface roughness around 3.9 nm, was used as graphene (pink) substrate and thermal absorber. **b**, Illustration of phase transformation from bilayer graphene/graphite to HD and CD under HPHT. The latter phase presumably possesses lower free energy. **c-f**, Optical evolution of FLG on Re substrate under HPHT. The 3L graphene (inset, optical image before transfer) was transferred onto Re surface and then compressed to 0.5 GPa at 300 K (**c**); Graphene/Re compressed to 32.0 GPa at 300 K (**d**); Graphene/Re heated to about 2924 K under 32.3 GPa (**e**); The sample quenched to ambient condition finally (**f**). Scale bars in **c-f** are 30 μm. The yellow dashed line highlights the morphology of graphene before laser heating. **g**, Raman spectra showing the phase evolution from the initial 3L graphene (exfoliation and after transferring) to the quenched 2D diamond synthesized after HPHT, identified by the characteristic $T_{2g}$ band. The spectra of quenched 2D diamond correspond to the positions marked as pink and violet in **f**. **h**, Cross-sectional HRTEM images of the 2D diamond. Low-magnification TEM image showing uniform morphology (top). HRTEM image shows the 2D diamond extent highlighted by green dashed lines, indicating a thickness of ~1.0 nm (bottom). **i**, EDS elemental mapping images of the 2D diamond cross-section for Au (top), C (middle) and Re (bottom) elements.

**Raman and PL characterizations**

Raman characterization serves as a robust tool for distinguishing diamond phase from other metastable structures. Photoluminescence (PL) analysis and microstructural characterization also provide complementary criteria for diamond identification, as discussed below. Besides, we further investigated the 2D diamond synthesis from graphene with varied thicknesses from several layers to several hundred nanometers (the details of synthesis conditions summarized in Supplementary Table 1). Our studies revealed the thickness-dependent transition conditions, that is, higher pressure and temperature are demanded for thinner precursors (Atomic structures and physical properties of the obtained 2D diamond will be discussed later).

The representative Raman and PL results of 2D diamond samples recovered from different synthesis conditions are shown in Figure 2. Laser heating method allows us to selectively treat the partial area of graphene precursor, while the rest remains opaque without direct laser irradiation (Figure 2a). After high pressure and room temperature (HPRT), the primary G and 2D peaks appeared at 1587 and 2774 cm$^{-1}$, arising from the doubly degenerate $E_{2g}$ mode and second-order scattering mode, respectively. As temperature increased, the enhanced D band at 1386 cm$^{-1}$ indicated that the structural distortion or transition occurred in graphene lattice, coinciding with the appearance of characteristic diamond Raman signal. Significantly, Raman intensity of diamond dramatically increased along the temperature gradient, while G band of graphene broadened and eventually disappeared. It is clear that $sp^3$ concentration in 2D diamond principally depends on the temperature under a certain pressure. The transparent flatten nanoflake recovered from 20.0 GPa and 1435 K exhibits rather low roughness (2.2 nm) and only a sharp Raman peak at 1332 cm$^{-1}$, and its fitted full width at half maximum (FWHM) of 3.6 cm$^{-1}$ approximates to commercial diamond crystal (3.3 cm$^{-1}$), suggesting the great crystalline quality of our formed 2D diamond (more results in Supplementary Fig. 4-5).

Furthermore, the quenched 2D diamond shows strong PL from the unique color centers, such as silicon-vacancy (SiV$^-$) and nitrogen-vacancy (NV$^0$). In Figure 2b, the intense and sharp peaks located at ~ 738 and ~ 575 nm correspond to the zero-phonon lines (ZPL) of SiV$^-$ and NV$^0$ centers, respectively, well consistent with literature results.[45,46] In addition, the temperature- and pressure-dependence of ZPL peak positions can principally prove the color center as well (Supplementary Fig. 6). These intense PL characters illustrate that the obtained 2D diamond has great application potential in quantum computing and sensor devices. Raman and PL mapping measurements enable us to effectively identify the diamond section, and the results demonstrate a non-uniform distribution of color centers in Figure 2c-e. The Si impurities originate from polydimethylsiloxane (PDMS) used during sample exfoliation process, and the N impurities may be attributed to the nitrogen-rich ambient air. The introduction of Si and N impurities serve as diamond nucleation sites and further lower the nucleation barriers.[32,47,48]

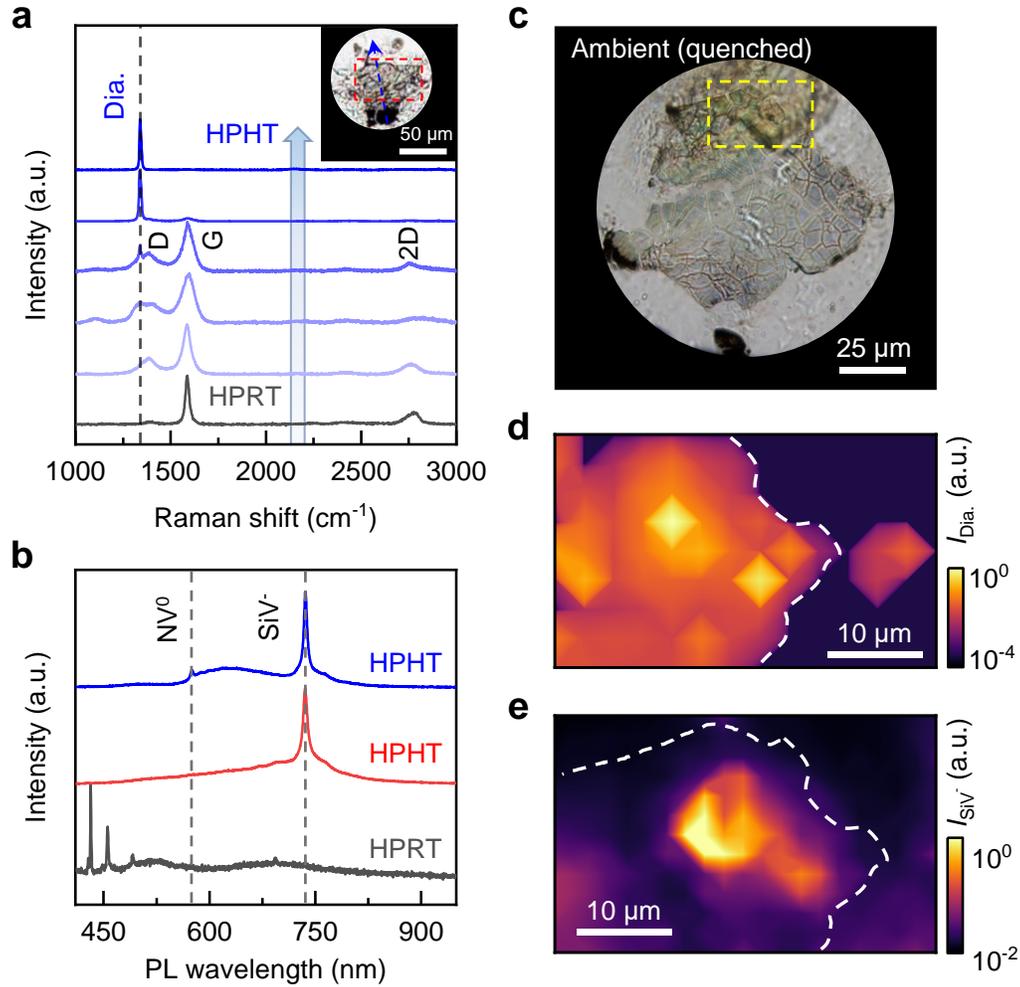

**Figure 2. Raman and PL measurements (405 nm excitation) of 2D diamond quenched from HPHT. a**, Raman spectra acquired from different positions along the blue dashed arrow in the optical image (inset). The partial graphene flake became transparent after 20.0 GPa and 1435 K, while the rest section remained opaque after HPRT treatment. The red box in inset indicated the preset laser heated region. Dia. means diamond. **b**, Typical PL spectra of recovered samples (blue and red) from HPHT with the remarkable SiV$^-$ (738 nm) and NV$^0$ (575 nm) color centers. The black line without any significant PL signal was taken from the opaque part in **a**. **c**, Optical image of quenched sample from 20.6 GPa and 1424 K to ambient condition for Raman and PL mapping measurements. **d**, 2D mapping result of diamond Raman intensity from the marked yellow box in **c**. **e**, Contour map of SiV$^-$ emission intensity of the recovered sample from the marked yellow box in **c**. The white dashed lines in **d** and **e** imply the boundaries between recovered sample and PTM. The slight variation of boundaries arises from positional displacement between two measurements, with the sample in **d** exhibiting an upward shift relative to **e**.

**Lattice structure and transition mechanism**

Atomic structure of 2D diamond and transition mechanism were investigated by HRTEM and selected area electron diffraction (SAED) (Figures 3 and Supplementary Fig. 7-10). HRTEM results reveal the coexistence of hexagonal diamond (Figures 3a-c and Supplementary Fig. 7e-f) and cubic diamond (Figures 3d-f and Supplementary Fig. 7g-i) when the samples were recovered from HPHT. A compact three-dimensional network structure displayed in Figure 3b corresponds to HD lattice observed from [001] zone axis. Carbon atoms are assembled as interconnected hexagonal rings in HD and the adjacent (100) layers with a $d$ spacing of 2.23 Å, consistent with the strong diffraction pattern. Alternatively, atomic arrangement in Figure 3c was observed from HD structure along [1$\bar{1}$0] zone axis as well. Figure 3e and 3f show the HRTEM images and SAED patterns of CD structures from [01$\bar{1}$] and [10$\bar{1}$] zone axes, respectively (more results in Supplementary Fig. 7). The measured $d$ spacing of ~ 2.12 Å is referred to the typical (111) plane.

It is worth noting that we observed RG structure next to the CD grain in Figure 3g-h, suggesting that RG may mediate HG to CD transition. The lattice planes with ~ 2.14 and 3.29 Å spacings are recognized as $(101)_{RG}$ and $(003)_{RG}$, respectively, and the interfacial angle between $(003)_{RG}$ is measured as ~ 78°. It is well known that the carbon layers in RG are stacked with ABC order, as highlighted in Figure 3h. The CD nuclei can be formed by sliding HG planes into ABC stacking, and then buckling RG basal planes into chair configuration.[35] The atomic arrangement indicates the orientational relations of $(002)_{HG} // (003)_{RG} // (111)_{CD}$ in our case. Besides, atomic structure of HG-HD interface was acquired as shown in Figure 3i. HG with interlayer distance of ~ 3.35 Å is adjacent to (100) plane of HD lattice with $d$ spacing of ~ 2.19 Å, which clearly suggests the orientation relation of $(002)_{HG} // (100)_{HD}$. The HG/RG/CD and HG/HD hybrid interfaces can be recovered from the transitional region with moderate temperature attributed to the energy distribution during laser heating and thermal diffusion effect. These atomic configurations of RG/CD and HG/HD interfaces agree well with the simulation results.[32,38]

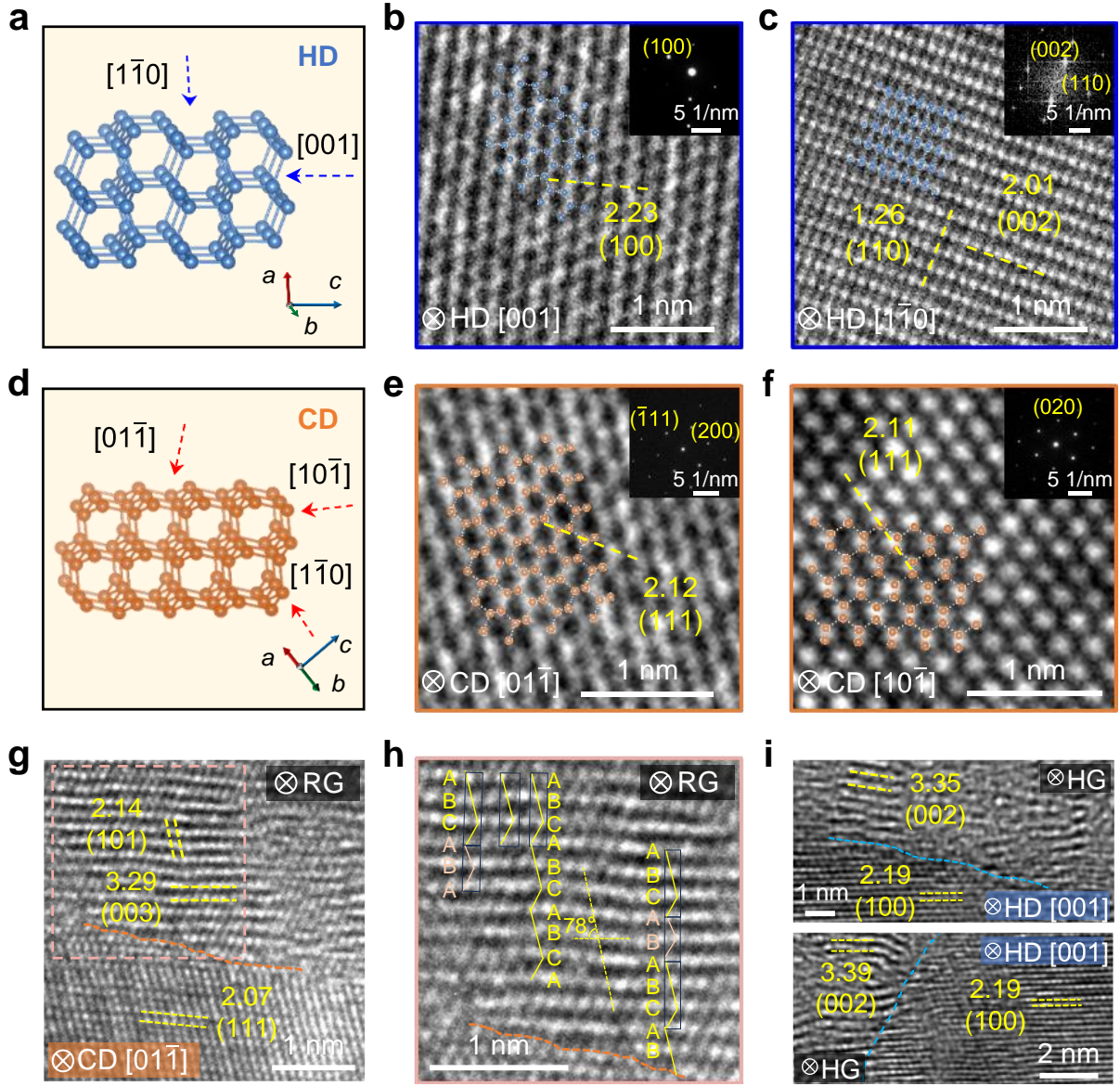

**Figure 3. Atomistic observations and transformation mechanism of the 2D diamond quenched from HPHT conditions. a**, Schematic of crystalline structure of HD. Atomic arrangements were observed along different zone axes (blue arrows). **b**, HRTEM image of HD from [001] zone axis. **c**, HRTEM image of HD from [1$\bar{1}$0] zone axis. The inset shows the fast Fourier transform (FFT) pattern. **d**, Schematic of lattice structure of CD. Atomic structures viewed along several zone axes (red arrows). **e,** HRTEM image of CD from [01$\bar{1}$] zone axis. **f**, HRTEM image of CD from [10$\bar{1}$] zone axis. The insets in **b** and **e-f** show the corresponding SAED patterns. **g**, Atomic arrangement of RG/CD hybrid interface. **h**, Details of ABC stacking orders selected from the marked orange box in **g**. **i**, Atomic arrangement of HG/HD hybrid interfaces. All labeled

numbers (unit: Å) mean the interplanar spacings measured from each HRTEM image. The 2D diamond samples for microstructural characterization were prepared with hundred-nanometer scale thickness.

Atomic nanostructures of the existent defects in 2D diamond can offer more insights into the transformation mechanism, including the apparent twin boundaries (TBs), stacking faults (SFs) and other diamond polytypes in Figure 4. The nanotwins in CD are found with sharing the typical {111} planes due to relatively lower defect energy[49,50] and further offsetting by step-like Σ3{211} TBs (Figure 4b and 4g). The Σ3{211} incoherent TBs are more inclined to maintain asymmetric structures with lower excess energy, and its length is determined by the spacing between neighboring {111} coherent TBs.[13] Figure 4d to 4f reveal the detailed fine structures of 2D diamond crystal along [01$\bar{1}$], [0$\bar{1}$1] zone axes, and CD nanotwin, respectively. The adjacent CD crystals share common (111) plane, as indicated by the yellow arrow in Figure 4f, and adopt the same atomic arrangement with mirror symmetry. Owing to the constraints imposed by coherent {111} TBs, the asymmetric Σ3(211) TBs are less mobile and may contribute to the potential strengthening of 2D diamond.[13]

Furthermore, diamond polytypes and the coherent graphene–2D diamond interfaces (termed "Gradia") shed light on the phase transition behavior under high pressure. The novel diamond polytypes surrounded by CD exhibit different lattice structures from HD and CD (Figure 4g). These diamond polytypes are coherently connected to CD lattice, and the rectangular pattern in FFT image illustrates a triple periodicity compared to CD, as shown in Figure 4h. These structural features can be potentially regarded as "M-diamond", reported with a range of non-CD polytypes previously.[11,12] All $d$-spacings of diamond polytypes extracted from FFT image in Figure 4h are summarized in Supplementary Table 2, in order to compare with the simulated $d$-spacings of diamond polytypes (such as 2H, 4H, 9R and 15R).[11,51] Besides, Figure 4i presents the identified interfacial structures between graphene precursor and the formed 2D diamond, primarily corresponding to the reported gradia-CA, gradia-CO and gradia-HC.[37]

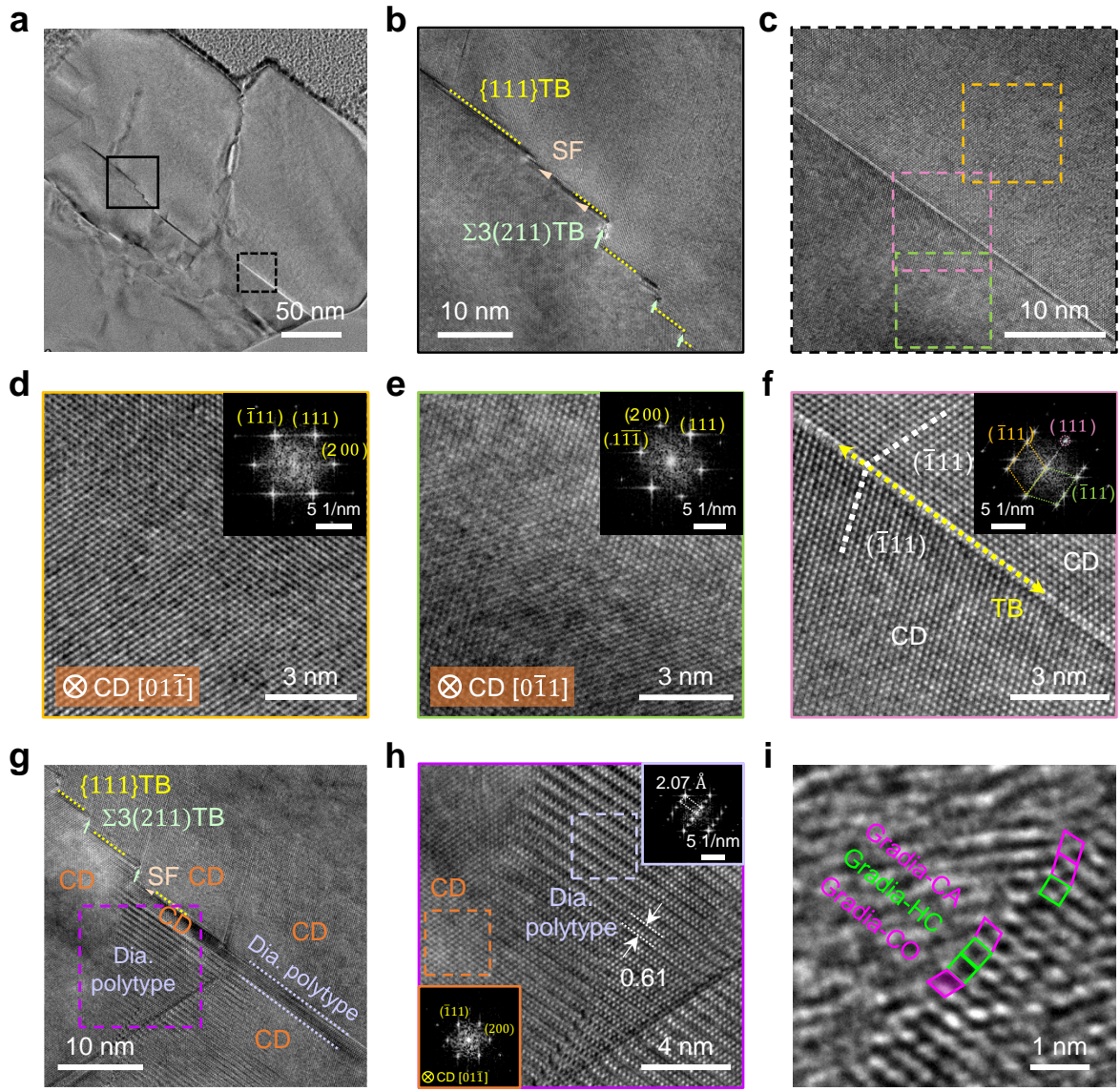

**Figure 4. Atomic characterizations of structural defects in the recovered 2D diamond. a**, A low-magnification TEM image of ultrathin diamond. **b**, High-magnification TEM image of the area selected from black solid box in **a**. The TBs (yellow dotted lines), SFs (arrowheads) and Σ3(211) TBs (pale green arrows) are well marked. **c**, High-magnification TEM image of the area selected from dotted box in **a**. **d-f**, HRTEM images of CD crystalline structures along [01$\bar{1}$] (**d**), [0$\bar{1}$1] (**e**) zone axes and TB (**f**, highlighted with yellow double arrow) from the regions marked in **c**. The insets show the corresponding FFT results. **g**, TEM image of diamond polytype coherently embedded in CD domains with TBs and SF. **h**, Magnified TEM image from the violet-boxed region in **g**. The FFT patterns of diamond polytype and CD are shown in insets as well. **i**, The identified

interfacial structure between graphene and 2D diamond. The 2D diamond samples for atomic characterization were prepared with hundred-nanometer scale thickness.

**Tunable bandgap and thermal stability of 2D diamond**

Electron energy loss spectroscopy (EELS) and absorption measurement were performed to explore intrinsic properties of ultrathin diamond. In Figure 5a, the strong 1s-σ* peak at 292.5 eV confirms the dominant $sp^3$ concentration, and the absence of 1s-π* peak at 283.2 eV represents the nearly complete phase transition from graphene precursor to 2D diamond. Our results indicate that $sp^3$ concentrations of 2D diamond vary from 71.3% to 89.9%. The optical bandgap of ultrathin diamond, determined with Tauc plot approach, sensitively depends on the $sp^3$ component of the recovered samples. The tunable bandgap energy of ultrathin diamond ranges from 1.4 to 1.9 eV as the $sp^3$ percentage increases (Figure 5b). The optical, Raman and PL results of the samples can be found in Supplementary Fig. 11. We further derived the Raman intensity ratio of diamond $T_{2g}$ ($I_{Dia}$) to graphene G band ($I_G$). As expected, the bandgap of 2D diamond exhibits a positive correlation with the calculated $I_{Dia}/I_G$, resulted from the increased $sp^3$ concentration, as shown in Figure 5c.

Moreover, thermal stability of ultrathin diamond was probed by thermal annealing treatment in inert argon atmosphere. Two recovered samples, labelled as #S1 and #S2 with different $I_{Dia}/I_G$ ratios, were progressively annealed up to 1000 °C for an hour. The evolution of $I_{Dia}/I_G$ is shown in Figure 5d-e (more details in Supplementary Fig. S12-S15). Obviously, the #S1 nanoflake with higher $sp^3$ percentage shows superior thermal stability up to 1000 °C, with the $I_{Dia}/I_G$ value (the average ratio ~ 11.8) almost unchanged. In contrast, the #S2 nanoflake with lower $sp^3$ percentage tends to graphitize at 900 °C as observed in Figures 5e and Supplementary Fig. 14. Raman intensity of $sp^3$-bonded phase at ~ 1340 cm$^{-1}$ drops gradually, accompanied by the enhanced G band of graphene. The obtained $I_{Dia}/I_G$ significantly decreased from 0.45 at 25 °C to 0.13 at 900 °C, and eventually vanished at 1000 °C (Figure 5f). Thermal stability of our prepared 2D diamond far surpasses that of many other diamond-like counterparts, such as diamond-like nanocomposite (DLN),[52] diamond-like carbon (DLC),[53] nano-diamond powder (Nano-dia.),[54] chemical vapor deposited diamond (CVD-dia.),[55] and Co-polycrystalline diamond (Co-PCD),[56] as summarized in Figure 5g.

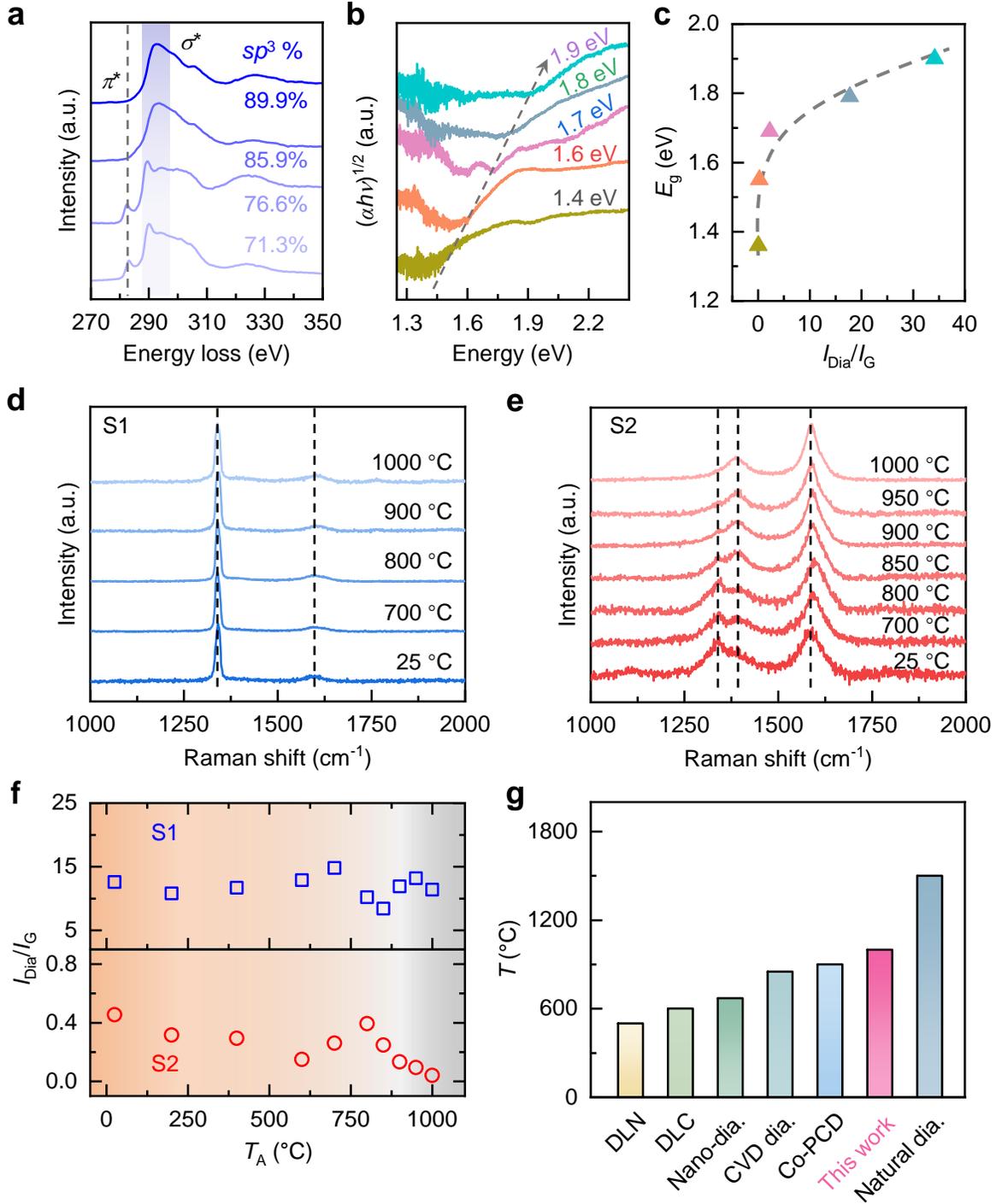

**Figure 5. The tunable bandgap energy and thermal stability of 2D diamond quenched from various synthesis conditions. a**, Carbon K-edge EELS results of the samples recovered from HPHT. The different curves indicate various temperature and pressure conditions. The features at 283.2 and 292.5 eV correspond to $sp^2$ and $sp^3$ hybridizations, respectively. **b**, Absorption spectra of ultrathin diamond with different $sp^3$ percentages. **c**, The $I_{Dia}/I_G$-dependent optical bandgaps of

quenched 2D diamond . The dashed line is used to guide the eyes. **d**, Raman spectra of ultrathin diamond (#S1) taken after different annealing temperatures. **e**, Annealing temperature-dependent Raman spectra of ultrathin diamond (#S2) from 25 to 1000 ºC. **f**, The evolution of $I_{Dia}/I_G$ of #S1 and #S2 with different annealing temperatures. **g**, Comparison of thermal stability of our 2D diamond with the reported carbon materials, including DLN,[52] DLC,[53] nano-dia.,[54] CVD-dia.,[55] Co-PCD[56] and natural diamond.[57] The 2D diamond samples for property investigations were prepared with hundred-nanometer scale thickness.

**DISCUSSION**

In conclusion, the quenchable 2D diamond with thicknesses down to three carbon layers has been successfully synthesized by the direct conversion of FLG under HPHT. Our observations confirm that 2D diamond exhibited well-crystalline $sp^3$ structures and strong PL peaks from SiV⁻ and NV⁰ color centers. Optical absorption and thermal annealing process indicated that the optical bandgaps and thermal stability of 2D diamond show $sp^3$ content-dependence. Atomic arrangements of 2D diamond revealed that intermediate RG mediate phase transition from HG to CD and the pathways along HG→RG→CD and HG→HD. This work can lay the foundation for the design and synthesis of novel carbon allotropes and further broaden their potential applications.

**EXPERIMENTAL METHODS**

**Two-dimensional diamond synthesis.** The graphene precursor with different thickness was mechanically exfoliated by PDMS. The exact layer number of FLG was determined by Raman spectroscopy together with atomic force microscopy. HPHT conditions can be realized via a homemade double-sided laser-heated DAC. The layout and photograph of our laser-heating system are illustrated in Supplementary Fig. 16. In short, two near-infrared lasers with tunable power were utilized to heat the sample, and the temperature can be determined by fitting the acuqired black-body radiation curve with Planck's theory. Importantly, the whole sytem should be calibrated with the standard lamp, such as the quartz tungsten halogen lamp (50 W, *Newport*) measured by *National Institute of Metrology of China* in our case. The high pressure was applied via a DAC with ~ 300 μm anvil culet. The sample chamber (~ 100 μm diameter) was prepared by drilling the pre-indented T301 stainless steel or Re gasket to form a through hole at the center. NaCl or KBr was served as both PTM and thermal insulator during HPHT experiments. Pressure was measured by the red shift of R1 fluorescence line of ruby ball. Considering the low absorbance of FLG, a Re foil with flat surface was utilized to play the role of graphene substrate and thermal absorber. The Re foil can be thinned to ~ 10 μm thick via a polishing machine and then shaped into small chips by laser drilling system. For the thick graphene precursor (normally > 50 nm), it can be sandwiched with two NaCl or KBr chips for the direct heating treatment. Once decompressed, the hydrophilic PTM can be conveniently removed by deionized water, and the recovered 2D diamond samples were transferred onto other desirable substrates for further investigations.

**Raman, PL, and absorption spectroscopy.** Raman and PL spectra of the recovered 2D diamond samples were measured using Horiba iHR550 spectrometer equipped with a CCD detector. The 405 nm laser was used as exciation source, and the beam size of focused laser approcached ~ 1 μm with a 100× objective. Raman signal of 2D diamond and PL spectrum of its color center were detected with 1800 and 150 $gmm^{-1}$ grating, respectively, while the pressure or temperature dependence of $SiV^-$ PL was investigated by 1800 $gmm^{-1}$ grating to ensure high spectral resolution. The prepared 2D diamond samples can be loaded into the homemade cryostat setup for *in situ* Raman and PL measurements, and the minimum temperature approximated 77 K realized via liquid nitrogen. Morevoer, the recovered 2D diamond samples were transferred onto the fused silica surface for transmission measurement. The absorption spectrum was collected by Horiba iHR550 spectrometer combined with halogen tungsten lamp with broad spectral range.

**HRTEM and EELS measurements.** In order to prepare HRTEM sample, the obtained 2D diamond was precoated with Pt layers, and then shapped into ultrathin slice with a thickness of 60-70 nm by focus ion beam (FEI Helios NanoLab 600i). HRTEM and SAED measurements were conducted by FEI Titan Cube 80-300 and JEM2100F with an acceleration voltage of 300 kV. Meanwhile, EELS spectra around the carbon K-edge were collected to investigate the bonding states in the recovered samples.

**Thermal stability characterizations.** The recovered 2D diamond samples were transferred onto the silica substrate for thermal annealing. Samples were annealed in tube furnace with flowing argon gas for 1 hour, and the annealing temperature varied from 200 to 1000 °C. Optical images and Raman mapping results were collected after annealing treatment under desired temperature.

**DATA AVAILABILITY**

The source data presented in this study have been deposited in the figshare database. All raw data generated during the current study are available from the corresponding authors upon request. Source data are provided with this paper.

## ASSOCIATED CONTENT

### Acknowledgements

This work was financially supported by the National Natural Science Foundation of China (Grants 52472040, 52072032, and 12090031), and the 173 JCJQ Program (Grant 2021-JCJQ-JJ-0159). Y.C. and J.L. thank the helpful discussions with Dr. L.X. Yang (HPSTAR) on the laser-heating system.

### Author contributions

Y.C. and J.L. conceived this research project and designed the experiments. J.L., G.D., J.F., and H.D. performed the HPHT synthesis of 2D diamond samples and their Raman and HRTEM characterizations. J.L., L.Z., W.X., S.C., and J.Y. attributed to the absorption measurements. J.L., W.H., P.Z., L.B., and Y.D. prepared the graphene samples on Re substrate. G.D., J.M., B.C., and J.Z. provided technical support for the construction of laser-heated DAC system. J.L. and Y.C. wrote the manuscript with the essential input of other co-authors. All authors have given approval of the final manuscript.

### Competing interest

The authors declare no competing financial interests.

### Supporting information

The Supporting Information is available free of charge online.

Detailed synthesis of 2D diamond under HPHT; characteristic Raman peak fitting and roughness analysis of the synthesized 2D diamond; temperature- and pressure-dependence of SiV$^-$ ZPL peak position; detailed microstructures of 2D diamond; optical and microstructural investigations of transient phase; Raman and PL spectra of samples used for absorption measurement; detailed transmittance evolution and Raman imaging of annealed samples; layout and photograph of the homemade double-sided laser heating apparatus; mechanical properties of 2D diamond and diamond-like samples.

**For Table of Contents only**

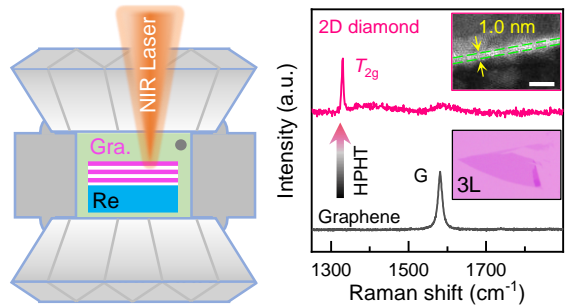

Two-dimensional diamond nanofilm with tuable properties synthesized from few-layer graphene under high pressure and high temperature conditions.